\documentclass[aps,prl,epsfig,showpacs,superscriptaddress,amsmath,amssymb,twocolumn,letterpaper]{revtex4-1}
\usepackage{amssymb}
\usepackage{amsmath}
\usepackage{latexsym}
\usepackage{graphicx}
\usepackage{dcolumn}
\usepackage{bm}
\usepackage{color}
\usepackage{soul}
\usepackage{bbold}

\newcommand{\be}{\begin{equation}}
\newcommand{\ee}{\end{equation}}
\newcommand{\ba}{\begin{eqnarray}}
\newcommand{\ea}{\end{eqnarray}}
\newcommand{\baa}{\begin{eqnarray*}}
\newcommand{\eaa}{\end{eqnarray*}}
\newcommand{\dg}{^{\dagger}}

\begin{document}

\title{Local Quantum Criticality of an Iron-Pnictide Tetrahedron}
\author{T.~Tzen Ong}
\affiliation{Center for Materials Theory, Department of Physics \& Astronomy, Rutgers University, Piscataway NJ 08854, USA}
\author{Piers Coleman}
\affiliation{Center for Materials Theory, Department of Physics \& Astronomy, Rutgers University, Piscataway NJ 08854, USA}
\affiliation{Department of Physics, Royal Holloway, University of London, Egham, Surrey TW20 0EX, UK}
\date{\today}

\begin{abstract}
Motivated by the close correlation between transition
temperature ($T_c$) and the tetrahedral bond angle of the As-Fe-As
layer observed in the iron-based superconductors,
we study the interplay between spin and orbital physics
of an isolated iron-arsenide tetrahedron embedded in a metallic
environment.  Whereas the spin Kondo effect is
suppressed to low temperatures by Hund's coupling, the orbital degrees of
freedom are expected to quantum mechanically quench at high
temperatures, giving rise to an overscreened, non-Fermi liquid ground-state.
Translated into a dense environment, this critical state
may play an important role in the superconductivity of these materials.
\end{abstract}
\maketitle

The discovery of superconductivity with $T_c = 26$K in LaFeAsO by Hosono {\it et. al.}
\cite{HosonoJACS08} has opened up a new field of iron-based high temperature
superconductors (SCs). Experimental studies suggest a range of electron
correlation effects.  For example, in the undoped parent
antiferromagnets, neutron scattering studies show
a small ordered moment in the pnictides,
but a significantly larger moment in the
chalcogenides\cite{DaiPhysC09}.  Similarly,
optical studies in $BaFe_2As_2$ and
$SrFe_2As_2$ indicate a spin-density wave (SDW) gap \cite{HuPRL08}
normally associated with itinerant electrons while
optical studies in LaFePO \cite{BasovNPhys09} and ARPES results in
$FeSe_{0.42}Te_{0.58}$ \cite{BaumbergerPRL10, HasanPRL09} suggest
significant correlation effects.
This diversity of behavior has prompted
an active debate on the role of electron correlation
effects. While some groups favor a more localized theoretical
description\cite{SiAbrahamsPRL08,HauleKotliarNPhys11}, many
theoretical treatments\cite{BasovChubukovNPhys11} have favored an itinerant, multi-band description.

One aspect of these systems that is poorly understood,
concerns the strong connection between the
many body physics and solid-state chemistry\cite{JohnstonAdvPhys10} .
In these superconductors the iron atom sits inside a tetrahedral cage of pnictide (As,P)
or chalcogenide (Te,Se) atoms. Experiments show that
the transition temperature is maximal when the bond-angles most closely approximate that
perfect tetrahedron ($\theta $(As-Fe-As) = 109.5$^{0}$)\cite{DaiNMat08,YamadaJPSJ08}. More recent studies show
\cite{Matsuda11} that when the tetrahedra are compressed along the
c-axis, the superconducting gap becomes anisotropic, and measurements also show the presence of line nodes \cite{MatsudaPRB10}.

\begin{figure}[bht!]
\begin{center}
\includegraphics[trim=0mm 0mm 0mm 0mm, clip, width=8cm]{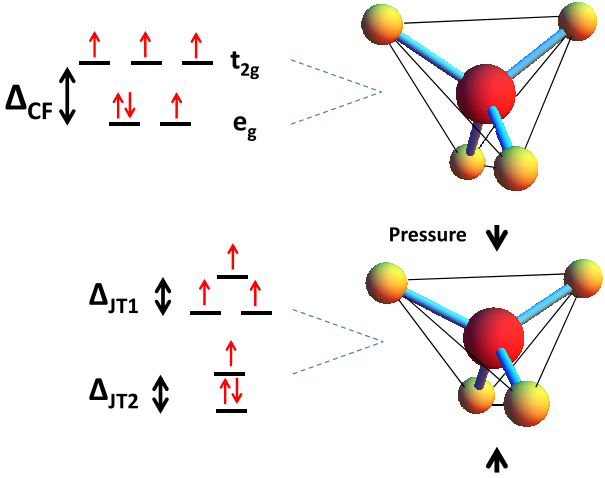}
\end{center}
\caption{Crystal field splitting of the $d$-electron orbitals in a
tetrahedron into the upper $t_{2g}$ and lower $e_g$ orbitals. The
lower diagram shows effect of compression along the c-axis,
which lifts the two-fold degeneracy in the $e_g$ orbitals.
}
\label{Fig:Tetrahedral diagram}
\end{figure}

Motivated by these issues, here
we develop a theory for an isolated  iron-tetrahedron
embedded within a conduction sea. One of our key observations, is that in addition to their spin
physics, the iron-based tetrahedra develop
an orbital degree of freedom
associated with the degenerate $e_{g}$ orbitals.
For conventional transition metal ions,
the Hund's coupling $J_H$ locks the unpaired
electrons together into a high spin configuration,
exponentially suppressing the spin-Kondo temperature to low
temperatures according to  an effect discovered by Schrieffer
\cite{SchriefferJAP67}. Here we show that unlike their spin counterparts, orbital fluctuations are not subject to the ``Schrieffer
effect'', giving rise to a unique situation in which the orbital
degrees of freedom behave as fluctuating quantum mechanical variables that result in an incoherent ``non-Fermi liquid'' ground-state. While departures from perfect tetragonality will
re-establish the Fermi liquid, a large temperature range of incoherent
metal behavior is expected to remain.
\begin{figure}[bht!]
\begin{center}
\includegraphics[trim=0mm 0mm 0mm 0mm, clip, width=8cm]{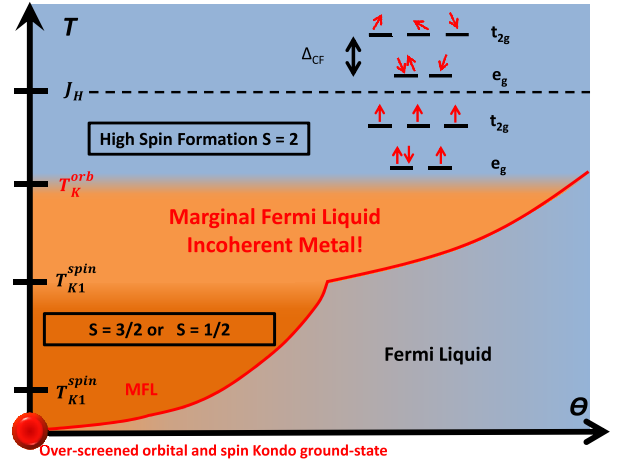}
\end{center}
\caption{Phase diagram of single Fe in an As tetrahedron. At $T \sim
J_H$, the Hund's coupling locks the spins of the Fe atom into an $S =
2$ state, and the orbital Kondo effect flows to strong coupling,
giving rise to a Marginal Fermi Liquid over-screened state at $T \sim
T_K^{orbital}$. The dangerously irrelevant spin Kondo interaction
finally drives the system to an over-screened spin Kondo ground state
at $T_{K2}^{spin}$. $\theta$ is the deviation away from perfect
tetrahedral symmetry, where $\theta_c = 109.5^o$, and will break the
$SU(2)^{orb}$ symmetry and restore Fermi liquid behaviour.}
\label{Fig:Schematic phase diagram}
\end{figure}
We assume that the FeAs tetrahedron contains an Fe$^{2+}$ ion in an
$3d^6$ configuration. In a tetrahedral environment, the five
$d$-electron orbitals are split by the crystal field
$\Delta_{CF}$ into three degenerate upper $t_{2g}$ orbitals and the
two degenerate lower $e_g$ orbitals, as shown in Fig. \ref{Fig:Tetrahedral diagram}.
When the tetrahedron is perfect, the unpaired electron in the lower $e_{g}$ orbitals can sit
in the $d_{3z^{2}-r^{2}}$ or the $d_{x^{2}-y^{2}}$ orbital, giving rise to an unquenched orbital degree of
freedom. Using a perturbative renormalization group treatment,
we  show that at high energies, orbital Kondo fluctuations dominate and the system flows to an
over-screened orbital Kondo fixed point. The residual low energy orbital degrees of freedom condense into
a single Majorana fermion that scatters electrons to produce a
Marginal Fermi liquid (MFL).  By analyzing the physics of this state using a
strong coupling expansion, we show that the subsequent low temperature
spin Kondo effect is also overscreened, so that the
the FeAs tetrahedron is a critical system down to the lowest temperatures.
The local physics of a single FeAs tetrahedron can be modelled by an Anderson model $H=H_{0}+H_{A}$
that describes the hybridization of Hund's coupled $d$-electrons in  five electron channels $\Gamma
\in [t_{2g}, e_g]$, where
\begin{eqnarray}\label{l}
H_{0}&=&\sum_{\vec{k} \sigma \Gamma} \epsilon_{\vec{k}} \, c_{\vec{k}
\Gamma \sigma}^{\dagger} c_{\vec{k} \Gamma \sigma}, \cr
H_{A} & = &
\sum_{\Gamma \in [t_{2g}, e_g]} (V_{\Gamma}
d_{\Gamma \sigma}^{\dagger} \psi_{\Gamma \sigma} + {\rm H.c.}
) +\epsilon_{\Gamma} \, d_{\Gamma \sigma}^{\dagger} d_{\Gamma
\sigma}
\nonumber
 \\
&+&
\sum_{\Gamma \in
[t_{2g}, e_g] }
+ U
\, n_{\Gamma \sigma} n_{\Gamma \overline{\sigma}} \, - \, J_H |\vec{S}|^2
\end{eqnarray}
Here, $\vec{S} = 1/2 \sum_{\Gamma } d_{\Gamma \alpha}^{\dagger} \vec{\sigma}_{\alpha \beta} d_{\Gamma \beta}$, is the total spin of the d-electrons, $\epsilon_{\Gamma}$ the  energy of the  crystal field level $\Gamma$, while $\psi \dg_{\Gamma\alpha }= \sum_{\vec{k}}c\dg_{k\Gamma\alpha}$ creates a conduction electron at the origin in channel $\Gamma$.
We shall assume that the Hubbard interaction, $U$, is the largest scale in the problem, with a hierarchy
of energy scales given by $U > D > J_H > V_{\Gamma}$, where $D$ is the electron bandwidth.

The effective low energy  Hamiltonian can be derived by a Schrieffer-Wolff
transformation, and is given by $H=H_{0}+H_{K}$, where
\begin{eqnarray}\label{KH2}
H_{K} & = &
\frac{J_2}{4S}
 \vec{S}\cdot \vec{\sigma}_{t_{2g}} (0)
 \nonumber \\
&  +& \frac{J}{2} \left( \frac{1}{2
S} \vec{S} \cdot \vec{\sigma}_{e_g} (0) + \frac{1}{2} \mathbb{1} \right)
\left( \vec{T} \cdot \vec{\tau} (0) + \frac{1}{2} \mathbb{1} \right)
\end{eqnarray}
where $\hat{\vec{\sigma}}_\Gamma (0) = \frac{1}{2}\psi_{\Gamma\alpha }^{\dagger}
\vec{\sigma}_{\alpha \beta } \psi_{\Gamma \beta }$ is the conduction electron
spin-density at the origin in channel $\Gamma$, while
\begin{equation}
\hat{\vec{\tau}} (0) =
\frac{1}{2}\sum_{\Gamma \Gamma' \in e_g} \psi_{\alpha \Gamma}^{\dagger}
\vec{\tau}_{\Gamma \Gamma'} \psi_{\alpha \Gamma'}
\label{eqn:KH3}
\end{equation}
describes the orbital moment of the conduction electrons in the
$e_{g}$ channel. The Kondo coupling in each channel is given by $J_{\Gamma} = 2|V_{\Gamma}|^2(1/|\epsilon_{\Gamma}| + 1/(U + \epsilon_{\Gamma}))$.

The first term in $H_{K}$ describes the Kondo interaction in the
$t_{2g}$ orbitals, while the second term, reminiscent of a local
``Kugel Khomskii'' interaction\cite{KugelSPU82, LeHurPRB04}, describes
the mixing of $e_{g}$ spin and orbital degrees freedom: for $S =
1/2$, this is the well-known SU(4) Kondo interaction, but for large
Hund's coupling, the system is locked into a high spin state with $S =
2$. Projected into this subspace, the interaction in the $e_g$
orbitals factorizes into spin and orbital Kondo interactions.

\begin{figure}[bht!]
\begin{center}
\includegraphics[angle=0,width=6cm]{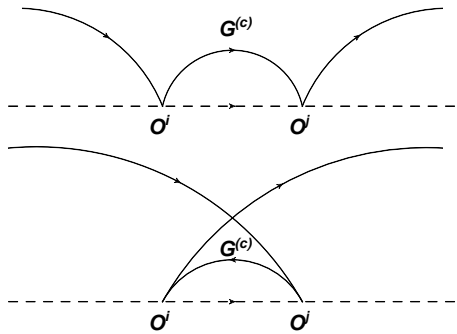}
\end{center}
\caption{The two one loop diagrams that contribute to the renormalization of $g_T$, $g_S$ and $g_m$. Here, the operator $O$ corresponds to the spin, iso-spin or ``mixed" Kondo operator, i.e. $\vec{O} \in [\vec{\sigma}, \vec{\tau}, \vec{\sigma} \vec{\tau}]$, and the dashed line corresponds to the local conduction electron propagator $G^{(c)}(\omega)$.}
\label{Fig:Perturbative RG diagrams}
\end{figure}

One of the effects of the projection into the high-spin manifold is an
$S$-fold reduction of the spin Kondo coupling, $J_S = J/4S $ but
strikingly, the the strength of the orbital Kondo interaction is
unaffected. This means that the Hund's
interaction will exponentially suppress the spin Kondo effect down to
a lower scale $T_K^{spin} / D \sim \left(T_K^{orb} /D \right)^{2S}$,
as in conventional transition metal ions, while leaving the orbital
Kondo effect unaffected.  This schism between the orbital and spin
Kondo effect drastically affects the physics, as we now confirm from a
renormalization analysis.
\ba
H_{K} & = &
  \frac{J}{2} \vec{T} \cdot \vec{\tau} + \frac{J}{4S} \vec{S} \cdot \vec{\sigma}_{e_g} \nonumber \\
      &    & + \frac{J}{2S} \left( \vec{S} \cdot \vec{\sigma}_{e_g} \right) \left( \vec{T} \cdot \vec{\tau} \right) + \frac{J_2}{4S} \vec{S}\cdot \vec{\sigma}_{t_{2g}}
      \label{Eq:KH4}
\ea
A perturbative RG treatment of Eq.~\ref{Eq:KH4} is carried out, with the Feynman
diagrams shown in Fig.~\ref{Fig:Perturbative RG diagrams}; and the
dimensionless coupling constants are defined as $g_S = \tfrac{J}{4 S}
\rho(E_F)$, $g_T = \tfrac{J}{2} \rho(E_F)$ and $g_m = \tfrac{J}{2S}
\rho(E_F)$. Since there are no orbital fluctuations in the $t_{2g}$
orbitals, we just obtain the standard Kondo result for $J_2$. In the $e_g$ channels, there are contributions to the
renormalization of $g_S$ and $g_T$ from the ``mixed" Kondo term, which
are proportional to $|\vec{T}|^2 g_m^2$ and $|\vec{S}|^2 g_m^2$
respectively, and the beta-functions are,
\ba
\beta_{g_T} & = & -2 g_T^2 - 2S(S+1) g_m^2 \nonumber \\
\beta_{g_S} & = & -2 g_S^2 - \frac{3}{2} g_m^2 \nonumber \\
\beta_{g_m} & = & 8 g_m^2 - 4 g_m g_T - 4 g_m g_S
\label{Eq:RG eqns 1}
\ea

Since both $g_S$ and $g_m$ are of order $O(1/S)$, while $g_T$ is of
order $O(1)$, clearly $\beta_{g_T}$ is of order $O(1)$ and
$\beta_{g_m} \sim O(1/S)$ and $\beta_{g_S} \sim O(1/S^2)$. Hence,
$g_T$ is the most relevant perturbation, and the system initially
flows towards the over-screened orbital fixed point at $T =
T_K^{orb}$.  The key result is that the quenching of the orbital
degrees freedom leaves behind an unscreened $S = 2$ spin moment.
coupled to five independent screening channels.

While orbital fluctuations are quenched below $T_{K}^{orb}$, we might expect that only remaining local
degrees of freedom would be the local $S = 2$ spin, Kondo coupled to the five
channels. From this point-of-view, the effective Hamiltonian will be
an overscreened five-channel $S=2$ Kondo model consisting of the first two terms in Eq.~\ref{Eq:5 channel Kondo
model} below.
\be
H = H_0 + \sum_{\Gamma} J_{\Gamma} \vec{S} \cdot \psi_{\Gamma
\alpha}^{\dagger} \frac{\vec{\sigma}_{\alpha \alpha'}}{2} \psi_{\Gamma
\alpha'} + H_{res}
\label{Eq:5 channel Kondo model}
\ee
However, this reasoning ignores certain subtle low-energy excitations generated by
the orbital Kondo effect, that we have anticipated by including a
further term $H_{res}$.  In fact, the orbital Kondo effect couples to
the spin up and spin down conduction electrons, and is thus
overscreened.  This gives rise to a well-documented MFL state with a zero-energy excitation described by a single
Majorana fermion $\phi(0)$~\cite{EmeryPRB92,ColemanPRB95,MaldacenaNucPhysB97}. The interaction with the spin and
conduction electrons must be taken into account.  This requires
decoupling the spin and orbital sectors of the $e_{g}$ conduction
electrons in terms of itinerant Majorana fermions, $(\psi^0,
\vec{\psi})$, and $(\chi^0, \vec{\chi})$, which respectively carry the
spin and orbital quantum numbers.

In the strong-coupling limit, the Hamiltonian is projected into the
degenerate orbital singlets subspace. Clearly, the spin Kondo term depends only on the $\psi$ fermions that carry spin, whereas the mixed Kondo term depends on both $\psi$ and $\chi$ fermions as it can mix the two singlet states. The Majorana fermion at the origin, $\phi(0) = \chi^1(0) \chi^2(0) \chi^3(0)$ is a three-fermion composite, and it couples to the $\chi$ fermions at the neighbouring site via the kinetic term; thus the first non-trivial term is $3^{rd}$-order, $\alpha = 3 t^3/4 g_T^2$. The effective strong coupling Hamiltonian, $\tilde{H}$, is therefore,
\ba \tilde{H} & = & it \sum_{n=1}^{n=\infty} \sum_{a = 0}^{3} \left[
\psi^a(n+1) \psi^a(n) + \chi^a(n+1) \chi^a(n) \right] \nonumber \\
              &   & + \alpha \phi(0) \chi^1(1) \chi^2(1) \chi^3(1) - i \frac{g_S}{2}
\vec{S} \cdot (\vec{\psi}(0) \times \vec{\psi}(0)) \nonumber \\
              &   & + \frac{g_m}{2} \psi^0(0) \vec{S} \cdot \vec{\psi}(0) \phi(0) \chi^0(0) + H_{0, t_{2g}} + H_{K, t_{2g}}
              \label{Eq:Strong coupling Hamiltonian}
\ea
where $H_{0,t_{2g}}$ and $H_{K, t_{2g}}$ are the kinetic energy of the conduction electrons and the Kondo interaction in the $t_{2g}$ channel. Dimensional analysis shows that the canonical dimensions of the Majorana fermions and coupling constants are $[\psi]_L = [\chi]_L = -\tfrac{1}{2}$, and $[g_S]_L = 0$ and $[g_m]_L
= \tfrac{1}{2}$. Therefore $g_m$ is irrelevant around the strong coupling fixed point.

We then carried out a one loop calculation of the renormalization of $g_S$. At $T = 0$, the local Majorana fermions propagators $G^{\chi}(\tau) = G^{\psi}(\tau) \sim \tfrac{1}{\tau}$, and
$G^{\phi}(\tau) = \tfrac{1}{2} sgn(\tau)$, which means that at one
loop order, there is a $(g_S/2)^2 ln \Lambda$ correction to
$g_s$, where $\Lambda$ is an UV cut-off, whereas $g_m$ does not contribute. This shows that the spin Kondo coupling,
$g_S$, is marginally relevant around the strong coupling fixed point, and will finally drive the system into an over-screened spin and orbital Kondo ground state with a different MFL. Therefore, the system is a quantum-critical incoherent metal down to 0 K.

\begin{figure}[thb!]
\begin{center}
\includegraphics[angle=0,width=8cm]{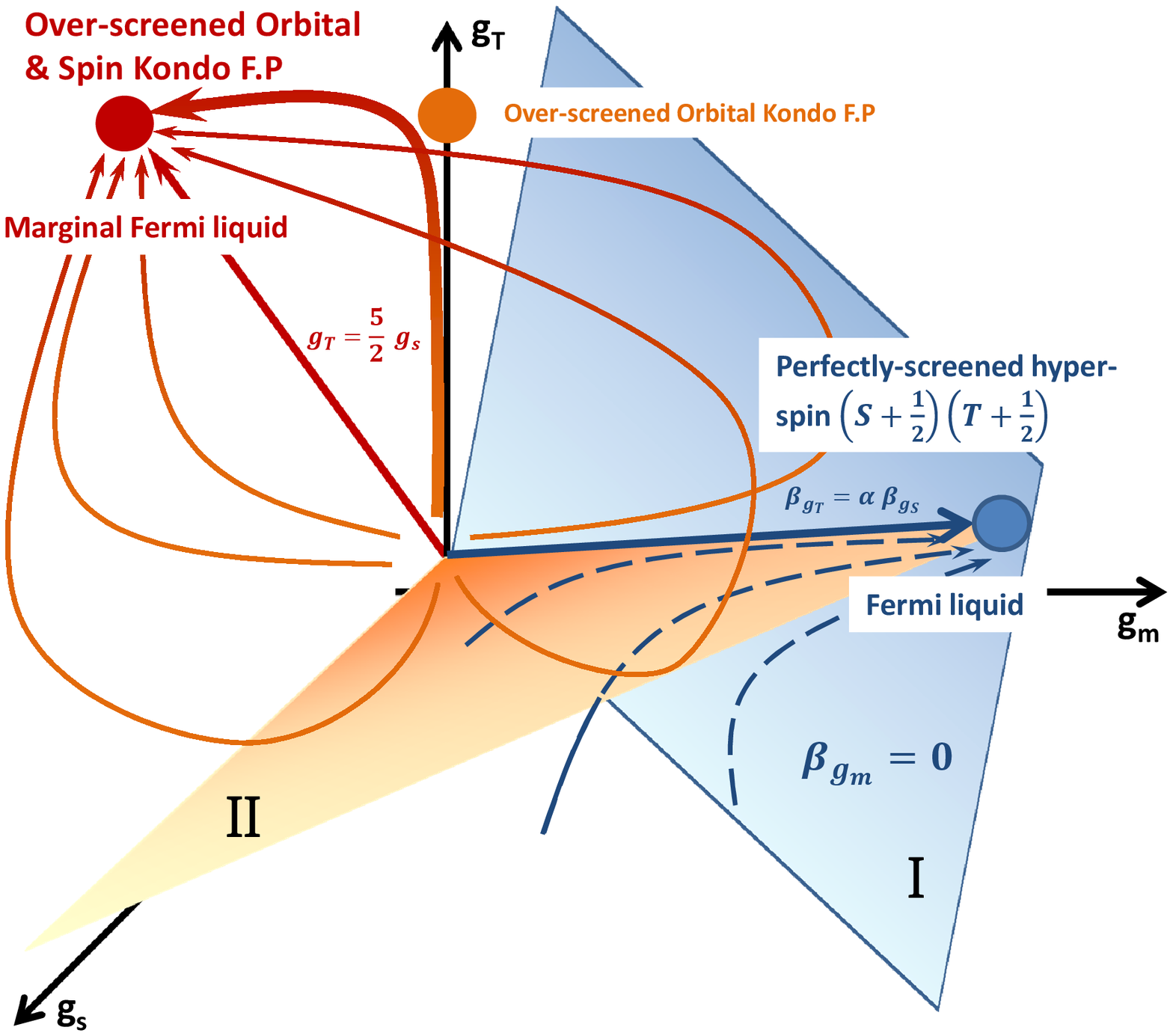}
\end{center}
\caption{Schematic plot of the RG flow. The perfectly screened hyper-spin Fermi liquid fixed point lies along the line $\beta_{g_T} = \alpha(S) g_S$ in plane I, and is stable against weak breaking of SU(4) symmetry \cite{LeHurPRB04}. For strong asymmetry, the system flows to the over-screened orbital and spin Kondo Marginal Fermi Liquid fixed point. The blue plane I is defined by $\beta_{g_m} = 0$, and the orange plane II is defined by $\beta_{g_T} = \alpha(g_m, S) \beta_{g_S}$. The bold line indicates the expected RG flow for a realistic FeAs tetrahedron.}
\label{Fig:RG flow gT gS}
\end{figure}

Fig.~\ref{Fig:RG flow gT gS} shows the renormalization flows for the system. The perfectly-screened Fermi liquid fixed point lies in plane I, defined by $\beta_{g_m} = 0$, along the line $\beta_{g_T} = \alpha(S) g_S$, where $\alpha(S = 2) = 12.2$. For $S = 1/2$, $\alpha(S) = 1$ and we recover the SU(4) Fermi liquid fixed point, which is characterized by perfect screening of the spin and orbital degrees of freedom.

The FeAs tetrahedron initially flows towards the over-screened orbital Kondo fixed point, where orbital fluctuations are quenched at $T_{K}^{orb}$. It will then cross-over to an over-screened orbital and spin Kondo fixed point at a lower temperature $T_K^{spin}$, and two different MFLs arise at $T_K^{orb}$ and $T_K^{spin}$ due to over-screened orbital and spin Kondo physics respectively. The over-screened spin Kondo fixed point with $n$ channels lies at $g^* = \tfrac{2}{n}(1 - \frac{2}{n} ln 2)$~\cite{GanPRL93}; hence, the over-screened orbital and spin fixed point lies on the line $g_T \approx \tfrac{n_s}{n_o} g_S \approx 5/2 g_S$.

Fig.~\ref{Fig:Schematic phase diagram} summarizes the phases, and the Fe impurity forms an $S = 2$ local
moment at a temperature $T \sim J_H$. Due to the Schrieffer mechanism for Kondo resonance narrowing, the
orbital and spin Kondo temperatures are split, $\left(T_K^{orb} / D
\right)^{2S} \sim T_K^{spin} / D$, and for $T_K^{orb}
< T < T_K^{spin}$, the system flows towards a two channel orbital Kondo fixed
point with MFL behaviour.

At temperatures below $T_K^{orb}$, the orbital fluctuations in the
$e_g$ orbitals are quenched leaving an $S = 2$ spin interacting with
five channels of conduction electrons. The system will first be screened from an $S = 2$ state to an
$S = \tfrac{1}{2}$ or $S = 1$ state, depending on the relative strengths of the $t_{2g}$ and $e_g$ Kondo couplings; and will finally be completely screened at a lower temperature $T_{K2}^{spin}$. It is possible
that these two scenarios could correspond to the chalcogenides and
pnictides respectively.

In practice, real systems deviate from perfect tetrahedral symmetry, and breaking of the SU(2)$^{orbital}$ symmetry would lift the $e_g$ orbitals degeneracy via a Jahn-Teller splitting $\Delta_{JT}$. It is well known from
two channel Kondo physics that this will restore Fermi liquid
behaviour below an energy scale $\Delta_{JT}^2/T_{K}^{orb}$. This
Fermi liquid behaviour will be seen only if $\Delta_{JT}^2 / T_{K}^{orb} >
T_{K1}^{spin}$. Raman spectroscopy indicates $\Delta_{JT} \sim 20 - 30$ meV \cite{CheongSSC10}, and $T_K \sim 0.1$ eV for some transition metals\cite{SteyertRMP68}. In general, we expect $T_{K}^{orb} > \Delta_{JT}$, meaning that the large $T_K^{orb}$ will protect the system against breaking of the tetrahedral symmetry, thus preserving the critical over-screened phases for a
range of applied pressures in experiments.

Various aspects of our theory could be tested in dilution experiments on materials such as $Ba(Ru_{1-x}Fe_xAs)_2$. The presence of local moment behaviour is expected to give rise to a Kondo resistance minimum. Our theory also predicts a logarithmic temperature dependence of the orbital susceptibility, $\chi_{o} \sim ln T$ below $T_{K}^{orb}$, which can be measured by resonant ultrasound spectroscopy, and a corresponding logarithmic divergence, $\chi_s \sim ln T$, of the spin susceptibility below $T_{K1}^{spin}$. In addition, due to the series of MFLs that arise at $T_{K}^{orb}$ and $T_{K2}^{spin}$, there would be a $C_v \sim T ln T$ behaviour down to $\Delta_{JT}^2/T_{K}^{orb}$.

The mechanisms for superconductivity and nematicity in the iron-based
SCs remain open issues. We observe that
critical orbital fluctuations within the $e_{g}$ orbitals
could provide a natural mechanism for nematic order.
It is also known from studies of the over-screened single Kondo
impurity\cite{EmeryPRB92, EmeryPRL93} that there
is a divergent susceptibility for composite SC pairing and it remains
an interesting question as to whether any of these divergent susceptibilities play a role in the dense materials.

We gratefully acknowledge discussions with Natan Andrei, Gabriel
Kotliar, Zlatko Tesanovic and Jan Zaanen on aspects of this work.
This work is supported by DOE grant DE-FG02-99ER45790.
\bibliographystyle{btxbst}

\end{document}